\documentclass[prd,aps,nopacs,floats,letterpaper,floatfix,nofootinbib]{revtex4}
\usepackage{amssymb,amsmath,amsbsy}
\usepackage[dvips]{graphicx}
\usepackage{dcolumn}
\usepackage[usenames]{color}
\usepackage{euscript}
\usepackage{mathrsfs}
\usepackage{enumerate}

\newcommand{\be}{\begin{equation}}
\newcommand{\ee}{\end{equation}}
\newcommand{\bel}[1]{\begin{equation}\label{#1}}
\newcommand{\ba}{\begin{eqnarray}}
\newcommand{\ea}{\end{eqnarray}}
\newcommand{\bal}[1]{\begin{eqnarray}\label{#1}}
\newcommand{\Nup}{$\dot{N}_{\rm max}$}
\newcommand{\Npl}{$\dot{N}_{\rm high}$}
\newcommand{\Nre}{$\dot{N}_{\rm re}$}
\newcommand{\Nlow}{$\dot{N}_{\rm low}$}

\newcommand{\Rup}{$R_{\rm max}$}
\newcommand{\Rpl}{$R_{\rm high}$}
\newcommand{\Rre}{$R_{\rm re}$}
\newcommand{\Rlow}{$R_{\rm low}$}
\newcommand{\Rmin}{$R_{\rm min}$}

\begin{document}

\title[Rates of Compact Binary Coalescences for Ground-based GW Detectors]{Predictions for the Rates of Compact Binary Coalescences Observable by Ground-based Gravitational-wave Detectors}

\author{J.~Abadie$^{29}$, 
B.~P.~Abbott$^{29}$, 
R.~Abbott$^{29}$, 
M,~Abernathy$^{66}$, 
T.~Accadia$^{27}$, 
F.~Acernese$^{19ac}$, 
C.~Adams$^{31}$, 
R.~Adhikari$^{29}$, 
P.~Ajith$^{29}$, 
B.~Allen$^{2,78}$
% ...
\\
(The LIGO Scientific Collaboration and the Virgo Collaboration)\\
and K.~Belczynski$^{80,81}$}
% ...
\address{$^{2}$Albert-Einstein-Institut, Max-Planck-Institut f\"ur Gravitationsphysik, D-30167 Hannover, Germany}
% ...
\address{$^{19}$INFN, sezione di Napoli $^a$; Universit\`a di Napoli 'Federico II'$^b$ Complesso Universitario di Monte S.Angelo, I-80126 Napoli; Universit\`a di Salerno, Fisciano, I-84084 Salerno$^c$, Italy}
% ...
\address{$^{27}$Laboratoire d'Annecy-le-Vieux de Physique des Particules (LAPP),  IN2P3/CNRS, Universit\'e de Savoie, F-74941 Annecy-le-Vieux, France}
% ...
\address{$^{29}$LIGO - California Institute of Technology, Pasadena, CA  91125, USA }
% ...
\address{$^{31}$LIGO - Livingston Observatory, Livingston, LA  70754, USA }
% ...
\address{$^{66}$University of Glasgow, Glasgow, G12 8QQ, United Kingdom }
% ...
\address{$^{78}$University of Wisconsin--Milwaukee, Milwaukee, WI  53201, USA }
% ...
\address{$^{80}$Los Alamos National Laboratory, CCS-2/ISR-1 Group, Los Alamos, NM, USA }
\address{$^{81}$Astronomical Observatory, University of Warsaw, Al.~Ujazdowskie 4, 00-478 Warsaw, Poland }
% ...

\date{March 25, 2010; DCC number: LIGO-P0900125}

\begin{abstract} 
We present an up-to-date, comprehensive summary of the rates for all types of compact binary coalescence sources detectable by the Initial and Advanced versions of the ground-based gravitational-wave detectors LIGO and Virgo.  Astrophysical estimates for compact-binary coalescence rates depend on a number of assumptions and unknown model parameters, and are still uncertain.  The most confident among these estimates are the rate predictions for coalescing binary neutron stars which are based on extrapolations from observed binary pulsars in our Galaxy.  These yield a likely coalescence rate of  $100$ Myr$^{-1}$ per Milky Way Equivalent Galaxy (MWEG), although the rate could plausibly range from $1$ Myr$^{-1}$ ${\rm MWEG}^{-1}$ to $1000$ Myr$^{-1}$ ${\rm MWEG}^{-1}$~\cite{Kalogera:2004tn}.  We convert coalescence rates into detection rates based on data from the LIGO S5 and Virgo VSR2 science runs and projected sensitivities for our Advanced detectors.  Using the detector sensitivities derived from these data, we find a likely detection rate of $0.02$ per year for Initial LIGO-Virgo interferometers, with a plausible range between $2\times10^{-4}$ and $0.2$ per year.  The likely binary neutron-star detection rate for the Advanced LIGO-Virgo network increases to $40$ events per year, with a range between $0.4$ and $400$ per year.
\end{abstract}

\maketitle

\section{Introduction}

The ground-based detectors LIGO, Virgo, and GEO 600 (see \cite{Abbott:2007kv, Sigg:2008, Acernese:2008, Grote:2008} for recent status reports) are rapidly improving in sensitivity.  The search of data from the last science run (LIGO S5, Virgo VSR1) of the Initial versions of these detectors is still ongoing (see \cite{Collaboration:2009tt, Abbott:2009qj} for upper limits on rates of low-mass binary mergers from the first part of the run).
By 2015, Advanced versions of these detectors should be taking data with a sensitivity approximately 10 times greater than the initial sensitivity, so that the detection volume will grow by a factor of about a thousand.  Such improvements in detector sensitivity mean that the first gravitational-wave signature of a compact-binary coalescence (CBC) event could be detected in the next few years.  

Theoretical predictions of astrophysical event rates represent a crucial input into the development and assessment of the detection process.  For example, Advanced LIGO can be tuned to increase its sensitivity in some frequency bands, and the relative event rates for different types of sources can aid the decision-making process for selecting the best detector configuration.  Additionally, as detector sensitivities improve, even upper limits will start to become astrophysically interesting.  They will begin to rule out the models that predict the highest detection rates, thereby allowing us to place stricter constraints on astrophysically interesting quantities such as compact-object natal kick velocities, the strength of massive-star winds, and the parameters of dynamically unstable mass-transfer processes in binary stars (e.g., accretion during the common-envelope phase) \cite{Grishchuk:2001, Belczynski:2008, MandelOShaughnessy:2010}.

The primary goal of this document is to provide an accessible, up-to-date, comprehensive summary of the rates of compact-binary coalescence sources, specifically those involving neutron stars (NSs), stellar-mass black holes (BHs) and intermediate-mass black holes (IMBHs).  This document aims to be a reference source for rate predictions for the gravitational-wave astrophysics community.  It can also provide an introduction to the literature on compact-binary coalescence rate estimates.   No new rate derivations are presented here, but we do provide a consistent conversion of merger rates into detection rates for the LIGO-Virgo network using the most up-to-date sensitivities of the Initial and Advanced LIGO
 detectors. 

Much work has been done in the field of predicting astrophysical rates for compact binary coalescences since classic papers by Phinney \cite{Phinney:1991ei} and Narayan, Piran \& Shemi \cite{Narayan:1991} appeared in 1991.  We do not attempt a thorough review of the entire body of literature on the subject.  Rather, we focus on a selection of papers representative of different approaches to rate prediction, emphasizing those papers which not only predict rates for CBCs, but also evaluate the systematic uncertainties in rate estimates.  We include the most recent papers from each group, and only those which appeared after 2000.  Additional background information can be found in the detailed review by Postnov \& Yungelson \cite{PostnovYungelson:2006}.  In particular, see Table 4 of \cite{PostnovYungelson:2006} and Tables 3 and 4 of Grishchuk et al.~\cite{Grishchuk:2001} for a partial list of historical CBC rate predictions.

New papers in the field are coming out at an ever-increasing pace, as better theoretical understanding allows more sophisticated models to be built, while additional electromagnetic observations of binaries with compact objects (pulsars and X-ray binaries) provide tighter constraints on those models (see, e.g., Kalogera et al.~\cite{2007PhR...442...75K}).  This version of the document is by necessity a snapshot of the field; only papers that have appeared {\it in print} by October 1, 2009 are included here.  However, this is meant to be a living document, which the authors will maintain in order to keep the information current.  

The document begins with an Executive Summary, which contains the coalescence rates for various CBC sources per Milky Way Equivalent Galaxy (MWEG), per $L_{10}$\footnote{$L_{10} \equiv 10^{10} L_{B,\odot}$, where $L_{B,\odot} = 2.16 \times 10^{33}$ erg/s is the blue solar luminosity \cite{LIGOS3S4Galaxies}.} or per Mpc$^3$ for NS-NS, NS-BH and BH-BH\footnote{BH-BH rates quoted in the Executive Summary do not include the contribution from dynamical interactions in dense stellar environments; see Section \ref{BHBHdyn} for details.} binaries, or per globular cluster (GC) for IMBH-IMBH binaries and intermediate-mass-ratio inspirals (IMRIs) into IMBHs.  Upper limit, plausible optimistic, likely, and plausible pessimistic rate estimates are given where available, all referenced to the existing literature.  Detection rates are also provided for fiducial values of the horizon distance (see Section \ref{ligoconversion} for definition) for both Initial and Advanced LIGO-Virgo networks.  Section \ref{ligoconversion} describes how rates per galaxy are converted into detection rates. Section \ref{sourcetypes} on individual sources provides a comprehensive list of currently available estimates in the published literature, with specific details on how each value was extracted from the literature.  A brief review of the methods by which these estimates were obtained is also included.
        
\section{Executive Summary \label{summary}}


At present, there are significant uncertainties in the astrophysical rate predictions for compact binary coalescences.  These arise from the small sample size of observed Galactic binary pulsars, from poor constraints for predictions based on population-synthesis models, and from the lack of confidence in a number of astrophysical parameters, such as the pulsar luminosity distribution.  The uncertainties in the coalescence rates, which can reach $\sim 1-2$ orders of magnitude in each direction from the most likely prediction, make it difficult to quote a single rate for a given source type.  Rather, for most sources, we suggest quoting a range of rates taken from Table \ref{galratesMWEG}  (for rates per Myr per Milky Way Equivalent Galaxy), Table \ref{galratesL10} (for rates per Myr per $L_{10}$), or Table \ref{galratesMpc3} (for rates per Myr per Mpc$^3$) as follows:
{\it plausible rate estimates for $\langle$merger type$\rangle$ mergers range from \Rlow\  to \Rpl\  with a likely rate estimate of around \Rre\ [citation from the Table]}.   Detection rates can be similarly quoted from Table \ref{detrates} for each generation of the LIGO-Virgo network; because the configuration of future detectors is not yet fully specified, it may be advisable to say that these detection rates were computed for a given horizon distance, provided in the first footnote to Table \ref{detrates}. 


\begin{table}[htb]\renewcommand{\arraystretch}{1.5}
\caption{Rate statement terminology.\label{terminology}}
\begin{tabular}{c@{\quad\vline\quad}l@{\quad\vline\quad}l}
\hline
Abbreviation & Rate statement & Physical significance\\
\hline
\Rup, \Nup \footnotemark[1] & Upper limit & Rates should be no higher than...\\
\Rpl, \Npl & Plausible optimistic estimate & Rates could reasonably be as high as...\\
\Rre, \Nre & Realistic estimate & Rates are likely to be...\\
\Rlow, \Nlow & Plausible pessimistic estimate & Rates could reasonably be as low as...\\
\hline
\end{tabular}
\footnotetext[1]{The symbols \Rup, \Rpl, etc., refer to rates per 
galaxy; the symbols \Nup, \Npl, etc., refer to detection rates.}
\end{table}

Where posterior probability density functions (PDFs) for rates are available, \Rre~refers to the PDF mean, \Rlow~and~\Rpl~are the $95\%$ pessimistic and optimistic confidence intervals, respectively, and \Rup~is the upper limit, quoted in the literature based on very basic limits set by other astrophysical knowledge (see Table \ref{terminology}).
However, many studies do not evaluate the rate predictions in that way, and for some speculative sources even estimates of uncertainties may not be available at present.  In these cases, we assign the rates estimates available in the literature to one of the four categories, as described in detail in section \ref{sourcetypes}.  The values in all tables in this summary Section are rounded to a single significant figure; in some cases, the rounding may have resulted in somewhat optimistic predictions.

\begin{table}[htb]   
\caption{Compact binary coalescence rates per Milky Way Equivalent 
Galaxy per Myr. \label{galratesMWEG}}
\begin{tabular}{c@{\quad\vline\quad}c@{\quad}c@{\quad}c@{\quad}c}
\hline
Source  & \Rlow & \Rre & \Rpl & \Rup\\
\hline

NS-NS (MWEG$^{-1}$ Myr$^{-1}$)
& 1~\cite{Kalogera:2004tn}\footnote{Lower end of 95\% 
confidence interval for the pulsar luminosity distribution yielding the 
lowest rate (Model 14) in Table 1 of \cite{Kalogera:2004tn}}
& 100~\cite{Kalogera:2004tn}\footnote{Peak rate for the 
reference pulsar luminosity distribution (Model 6) in Table 1 of 
\cite{Kalogera:2004tn}}
& 1000~\cite{Kalogera:2004tn}\footnote{Upper end of 95\% 
confidence interval for the pulsar luminosity distribution yielding the 
highest rate (Model 15) in Table 1 of \cite{Kalogera:2004tn}}
& 4000~\cite{Kim:2006}\footnote{Mean rates plus $2 \sigma$ for Type 
Ib/Ic supernova~\cite{Kim:2006}, values from \cite{Cappellaro:1999} }\\

NS-BH   (MWEG$^{-1}$ Myr$^{-1}$)
& 0.05~\cite{Oshaughnessy:2008}\footnote{The left edge of the
probability distribution peak for NS-BH in Figure 6 of
\cite{Oshaughnessy:2008}} & 3~\cite{Oshaughnessy:2008}\footnote{The center of the probability 
distribution peak for NS-BH in Figure 6 of \cite{Oshaughnessy:2008}}
& 100~\cite{Oshaughnessy:2008}\footnote{The right edge of the 
probability distribution peak for NS-BH in Figure 6 of 
\cite{Oshaughnessy:2008}}
&\\

BH-BH (MWEG$^{-1}$ Myr$^{-1}$)
& 0.01~\cite{2007PhR...442...75K}\footnote{The left edge of the 
probability distribution peak for BH-BH in Figure 15 of 
\cite{2007PhR...442...75K}}
& 0.4~\cite{2007PhR...442...75K}\footnote{The center of the probability 
distribution peak for BH-BH in Figure 15 of \cite{2007PhR...442...75K}}
&  30~\cite{2007PhR...442...75K}\footnote{The right edge of the 
probability distribution peak for BH-BH in Figure 15 of 
\cite{2007PhR...442...75K}}
& \\  

IMRI into IMBH (GC$^{-1}$ Gyr$^{-1}$)
&  &  & 3~\cite{Mandel:2007rates}\footnote{Estimate from binary 
hardening via three-body interactions assuming the inspiraling object is 
a neutron star (Section 2.1 of \cite{Mandel:2007rates})} 
& 20~\cite{Mandel:2007rates}\footnote{Upper limit of $300 M_\odot/m$ per 
$10^{10}$ years per cluster (Section 3.3 of \cite{Mandel:2007rates}), 
assuming the inspiraling object $m=1.4\ M_\odot$ is a neutron star} \\

IMBH-IMBH (GC$^{-1}$ Gyr$^{-1}$)
& & & 0.007~\cite{imbhlisa-2006}\footnote{Assumes that $10\%$ of all 
globular clusters are sufficiently massive and have a sufficient binary 
fraction to form an IMBH-IMBH binary once in their lifetime, taken to be 
13.8 Gyr~\cite{imbhlisa-2006}}
& 0.07~\cite{imbhlisa-2006}\footnote{Assumes that all globular clusters 
are sufficiently massive and have a sufficient binary fraction to form 
an IMBH-IMBH binary once in their lifetime, taken to be 13.8 
Gyr~\cite{imbhlisa-2006}}
\\
\hline
\end{tabular}
\end{table}

In the simplest models, the coalescence rates are assumed to be proportional to the stellar birth rate in nearby spiral galaxies, which can be estimated from their blue luminosity\footnote{Blue-light luminosity may not be a perfect tracer of current star-formation rate (see, e.g., Kennicutt et al.~\cite{Kennicutt:2009}); however, it was useful for scaling the observations of early interferometers because it allowed Kopparapu et al.~\cite{LIGOS3S4Galaxies} to compile a galaxy catalog that is relatively complete out to $\lesssim 30$ Mpc.}.   We therefore express the coalescence rates per unit $L_{10}$ (i.e., $10^{10}$ times the Solar blue-light luminosity $L_{B,\odot}$) in Table \ref{galratesL10}, using the conversion factor of $1.7\ L_{10}/$MWEG \cite{Kalogera:2000dz}.  There is a danger to using blue-light luminosity as a conversion factor, however: although blue-light luminosity is a reasonable indication of the current star-formation rate in spiral galaxies, it does not accurately track star-formation rates in the past.  In particular, scaling to blue-light luminosity ignores the contribution of older populations in elliptical galaxies \cite{pacheco:2005}.  In the future, when the contribution of elliptical galaxies is properly included in published studies, merger rates will be more naturally quoted per unit volume, rather than per MWEG or per $L_{10}$.  We therefore include rates per Mpc$^3$ in Table \ref{galratesMpc3}; however, this table still includes only the contribution from spiral galaxies like the Milky Way, using the conversion factor $0.0198\ L_{10}/$Mpc$^3$ from Section 3.1 of Kopparapu et al.~\cite{LIGOS3S4Galaxies}.

\begin{table}[htb]   
\caption{Compact binary coalescence rates per $L_{10}$ per Myr.\protect\footnote{See footnotes in Table \ref{galratesMWEG} for details on the sources of the values in this Table} \label{galratesL10}}
\begin{tabular}{c@{\quad\vline\quad}c@{\quad}c@{\quad}c@{\quad}c}
\hline
Source & \Rlow & \Rre & \Rpl & \Rup\\
\hline

NS-NS ($L_{10}^{-1}$ Myr$^{-1}$)
& 0.6
& 60~\cite{Kalogera:2004tn}
& 600~\cite{Kalogera:2004tn}
& 2000~\cite{Kim:2006}
\\

NS-BH   ($L_{10}^{-1}$ Myr$^{-1}$)
& 0.03~\cite{Oshaughnessy:2008}
& 2~\cite{Oshaughnessy:2008}
& 60~\cite{Oshaughnessy:2008}
&\\

BH-BH  ($L_{10}^{-1}$ Myr$^{-1}$)
& 0.006~\cite{2007PhR...442...75K}
& 0.2~\cite{2007PhR...442...75K}
& 20~\cite{2007PhR...442...75K}
& \\  
\hline
\end{tabular}
\end{table}

\begin{table}[htb]   
\caption{Compact binary coalescence rates per Mpc$^3$ per Myr.\protect\footnote{See footnotes in Table \ref{galratesMWEG} for details on the sources of the values in this Table} \label{galratesMpc3}}
\begin{tabular}{c@{\quad\vline\quad}c@{\quad}c@{\quad}c@{\quad}c}
\hline
Source & \Rlow & \Rre & \Rpl & \Rup\\
\hline

NS-NS (Mpc$^{-3}$ Myr$^{-1}$)
& 0.01~\cite{Kalogera:2004tn}
& 1~\cite{Kalogera:2004tn}
& 10~\cite{Kalogera:2004tn}
& 50~\cite{Kim:2006}
\\

NS-BH   (Mpc$^{-3}$ Myr$^{-1}$)
& $6\times10^{-4}$~\cite{Oshaughnessy:2008}
& 0.03~\cite{Oshaughnessy:2008}
& 1~\cite{Oshaughnessy:2008}
&\\

BH-BH  (Mpc$^{-3}$ Myr$^{-1}$)
& $1\times10^{-4}$~\cite{2007PhR...442...75K}
& 0.005~\cite{2007PhR...442...75K}
& 0.3~\cite{2007PhR...442...75K}
& \\  
\hline
\end{tabular}
\end{table}

\begin{table}[htb]   
\caption{\label{detrates} Detection rates for compact binary coalescence sources. }
\begin{tabular}{c@{\quad}c@{\quad\vline\quad}c@{\quad}c@{\quad}c@{\quad}c}
\hline
IFO & Source\footnotemark[1] & \Nlow & \Nre & \Npl & \Nup\\
& & yr$^{-1}$ & yr$^{-1}$ & yr$^{-1}$ & yr$^{-1}$\\
\hline
& NS-NS 
& $2\times10^{-4}$ & 0.02 & 0.2 & 0.6 \\
& NS-BH & $7\times10^{-5}$  & 0.004 & 0.1 & \\
Initial & BH-BH & $2\times10^{-4}$ & 0.007 & 0.5  & \\
& IMRI into IMBH &  &  & $<0.001$\footnotemark[2] & 0.01\footnotemark[3] \\
& IMBH-IMBH & &  & $10^{-4}$\footnotemark[4]  & $10^{-3}$\footnotemark[5] \\
\hline
& NS-NS 
& 0.4 & 40 & 400 & 1000\\
& NS-BH  & 0.2  & 10 & 300 & \\
Advanced & BH-BH & 0.4 & 20 & 1000 & \\
& IMRI into IMBH &  & & 10\footnotemark[2] & 300\footnotemark[3] \\
& IMBH-IMBH & & & 0.1\footnotemark[4] & 1\footnotemark[5]\\
\hline
\end{tabular}

\footnotetext[1]{To convert the rates per MWEG in Table \ref{galratesMWEG} 
into detection rates, optimal horizon distances of 33 Mpc / 445 Mpc are 
assumed for NS-NS inspirals in the Initial / Advanced LIGO-Virgo networks.  For 
NS-BH inspirals, horizon distances of 70 Mpc / 927 Mpc are 
assumed.  For BH-BH inspirals, horizon distances of 161 Mpc / 
2187 Mpc are assumed.  These distances correspond to a choice of $1.4\ 
M_\odot$ for NS mass and $10\ M_\odot$ for BH mass.  Rates for IMRIs 
into IMBHs and IMBH-IMBH coalescences are quoted directly from the 
relevant papers without conversion.  See Section \ref{ligoconversion} 
for more details.}

\footnotetext[2]{Rate taken from the estimate of BH-IMBH IMRI rates 
quoted in \cite{Mandel:2007rates} for the scenario of BH-IMBH binary 
hardening via 3-body interactions; the fraction of globular clusters 
containing suitable IMBHs is taken to be $10\%$, and no interferometer 
optimizations are assumed.}

\footnotetext[3]{Rate taken from the optimistic upper limit rate quoted 
in \cite{Mandel:2007rates} with the assumption that all globular 
clusters contain suitable IMBHs; for the Advanced network rate, the 
interferometer is assumed to be optimized for IMRI detections.}

\footnotetext[4]{Rate taken from the estimate of IMBH-IMBH ringdown 
rates quoted in \cite{imbhlisa-2006} assuming 10\% of all young star 
clusters have sufficient mass, a sufficiently high binary fraction, and 
a short enough core collapse time to form a pair of IMBHs.}

\footnotetext[5]{Rate taken from the estimate of IMBH-IMBH ringdown 
rates quoted in \cite{imbhlisa-2006} assuming all young star clusters 
have sufficient mass, a sufficiently high binary fraction, and a short 
enough core collapse time to form a pair of IMBHs.}

\end{table}


\section{Conversion from Merger Rates to Detection Rates\label{ligoconversion}}

Although some publications quote detection rates for Initial and Advanced LIGO-Virgo networks directly, the conversion from coalescence rates per galaxy to detection rates is not consistent across all publications.   Therefore, we choose to re-compute the detection rates as follows.\footnote{Rates of IMRIs into IMBHs and IMBH-IMBH coalescences are an exception: because of the many assumptions involved in converting rates per globular cluster into LIGO-Virgo detection rates, we quote detection rates for these sources directly as they appear in the relevant publications.}

The actual detection threshold for a network of interferometers will depend on a number of factors, including the network configuration (the relative locations, orientations, and noise power spectral densities of the detectors), the characteristics of the detector noise (its Gaussianity and stationarity), and the search strategy used (coincident vs. coherent search) (see, e.g., \cite{Fairhurst:2007qj}).  A full treatment of these effects is beyond the scope of this paper.  Instead, we estimate event rates detectable by the LIGO-Virgo network by scaling to an average volume within which a single detector is sensitive to CBCs above a fiducial signal-to-noise ratio (SNR) threshold of $8$.  This is a conservative choice if the detector noise is Gaussian and stationary and if there are two or more detectors operating in coincidence.\footnote{The real detection range of the network is a function of the data quality and the detection pipeline, and can only be obtained empirically.  However, we can argue that our choice is not unreasonable as follows.  We compute below that the NS-NS horizon distance for the Initial-era interferometers is $D_{\rm horizon}=33$ Mpc.  According to Eq.~(\ref{NG}), this corresponds to an accessible volume of $\sim 150$ MWEGs or $\sim 250$ $L_{10}$.  Meanwhile, the $90\%$-confidence upper limit on NS-NS rates from a year and a half of data (including approximately half a year of double-concident data between Hanford and Livingston interferometers) in the fifth LIGO science run was $1.4\times10^{-2}$ yr$^{-1}$ $L_{10}^{-1}$ \cite{Abbott:2009qj}.  The probability of observing zero events is $10\%$ with a Poisson distribution with mean $2.3$, so this $90\%$-confidence upper limit corresponds to a detection volume of $\sim 160$ $L_{10}$ in $1.5$ years, which is somewhat less than our estimate of $250$ $L_{10}$.  However, this was the detection volume obtained with two LIGO 4-km detectors and one 2-km detector, H2, co-located with H1; the detection volume should be increased when using the two 4-km LIGO detectors and the Virgo detector in a network, yielding a better match with the estimated volume.}  Event rates for searches with an optimal horizon distance (see below) greater than the local over-density ($\gtrsim 30$ Mpc) but less than the scale on which cosmological effects become significant ($\lesssim 1$ Gpc) scale linearly with the sensitive volume and thus the cube of the inverse SNR threshold.  Readers can, therefore, appropriately adjust the quoted rates when considering detector networks with different sensitivities.

The detection rate for a given CBC type in a LIGO-Virgo search is equal to
\begin{equation}
\dot{N}=R \times N_G,
\end{equation}
where $R$ is the coalescence rate of that type of binary per galaxy  (given in Table \ref{galratesMWEG}) and $N_G$ is the number of galaxies accessible with a search for the relevant binary type. 

The reach of a given search is characterized by the optimal horizon distance 
$D_{\rm horizon}$.  For a single detector, the optimal horizon distance is defined as that distance at which an optimally-oriented, overhead source can be detected with a signal-to-noise ratio (SNR) $\rho(D_{\rm horizon})=8$, where
\bel{SNR}
\rho=\sqrt{4\int_0^{f_{\rm ISCO}} \frac{|\tilde{h}(f)|^2}{S_n(f)}df},
\ee
$S_n(f)$ is the noise power spectral density (PSD), $|\tilde{h}(f)|$ is the 
frequency-domain waveform amplitude
\bel{tildeh}
|\tilde{h}(f)|=\frac{2 c}{D} \left(\frac{5 G \mu} {96 c^3}\right)^{1/2} 
	\left(\frac{G M}{\pi^2 c^3}\right)^{1/3} f^{-7/6},
\ee
$D$ is the luminosity distance to the source, $M$ is the total mass, $\mu$ is the reduced mass, 
and the frequency $f_{\rm ISCO}$ of the innermost stable circular orbit (ISCO) is set to $f_{\rm ISCO}=c^3/(\pi\ 6^{1.5}\ G\ M) \sim 4396 / (M/M_\odot)$ Hz.  This calculation is a conservative estimate of the SNR, since it includes the inspiral portion of the waveform only, and ignores the merger and ringdown, which will not contribute significantly to the SNR for low-mass binaries.  Note that these expressions do not include the cosmological redshift, which is neglected in this document\footnote{The effect of cosmological redshift will only be significant for
BH-BH binaries in Advanced LIGO, where the horizon distance will correspond to a redshift of $z \sim 0.4$.  Redshift scales the masses in Eq.~(\ref{tildeh}) by a factor of $1+z$, thus increasing the waveform
amplitude, but decreases the ISCO frequency in Eq.~(\ref{SNR}) by the same factor, thus reducing the bandwidth of low-frequency signals.  These competing effects are further compounded by the redshift dependence of astrophysical of quantities like metallicities and star formation rates \cite{OShaughnessy:2008sfr}.}. 

\begin{figure}[htb]
\centering
\includegraphics[keepaspectratio=true, width=6in]{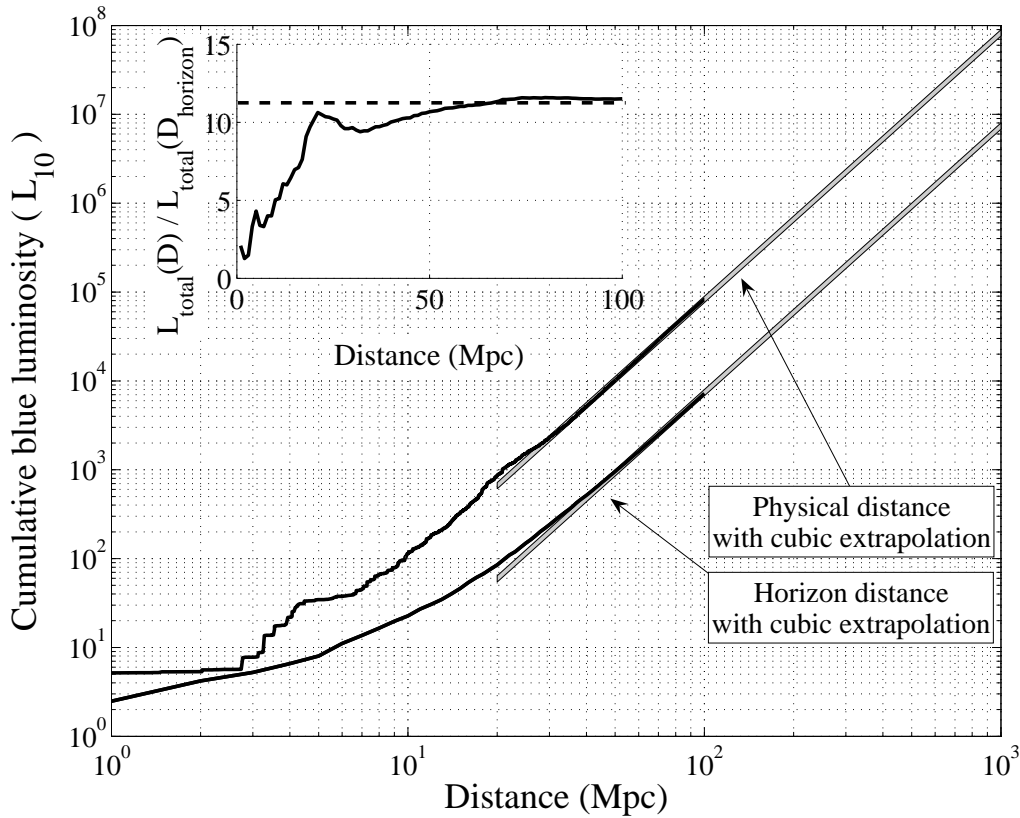}
\caption{The total blue-light luminosity within a sphere of a given radius (top curve) and the accessible blue-light luminosity for a given horizon distance $D_{\rm horizon}$, taking location and orientation averaging into account (bottom curve).   Gray shaded lines are cubic extrapolations.  The inset shows the ratio between the top and bottom curves, which asymptotes to $2.26^3$, as discussed in the text.  Reproduced from \cite{LIGOS3S4Galaxies} by permission of the AAS.}
\label{fig:L10vD}
\end{figure}

The number $N_G$ of galaxies accessible with that particular search is a function of the horizon  distance, $D_{\rm horizon}$.  The bottom curve of Figure \ref{fig:L10vD}, reproduced from \cite{LIGOS3S4Galaxies}, shows the amount $C_L$ of accessible blue-light luminosity in units of $L_{10}$ as a function of $D_{\rm horizon}$.\footnote{See Section \ref{summary} for a discussion of the limitations of scaling merger rates to blue-light luminosity.}  To convert this blue-light luminosity into the number $N_G$ of accessible Milky Way Equivalent Galaxies (MWEGs), a conversion factor of
\begin{equation}
N_G\ ({\rm MWEGs}) = 1.7\ C_L\  (L_{10})
\end{equation} should be used.  (This conversion factor follows from the discussion in \cite{Kalogera:2000dz}).

The following formula is a good approximation to $N_G(D_{\rm horizon})$ 
once the local over-density is averaged out at larger distances $D_{\rm 
horizon}\gtrsim 30$ Mpc:
\begin{equation}\label{NG}
N_G=\frac{4}{3}\pi \left(\frac{D_{\rm horizon}}{\rm Mpc}\right)^3 (2.26)^{-3} 
	(0.0116).
\end{equation}
Here, $1/2.26$ is the correction factor included to average over all sky locations and
orientations\footnote{The so-called ``sensemon range'' is the radius of a sphere whose volume is equal
to the volume in which an interferometer could detect a source at $\rho \ge 8$, taking
all possible sky locations and orientations into account.  The factor
of $1/2.26$ for the ratio between the optimal horizon distance and the
sensemon range is computed in \cite{FinnChernoff:1993}.  This factor neglects cosmological redshift corrections, which could be included using the framework described in \cite{Finn:1996}.} \cite{FinnChernoff:1993}, and $1.16\times 10^{-2}$ Mpc$^{-3}$ is the extrapolated density of MWEGs in space \cite{LIGOS3S4Galaxies}.  We use Eq.~(\ref{NG}) for all rate conversions in this document.

\begin{figure}[htb]
\centering
\includegraphics[keepaspectratio=true, width=6in]{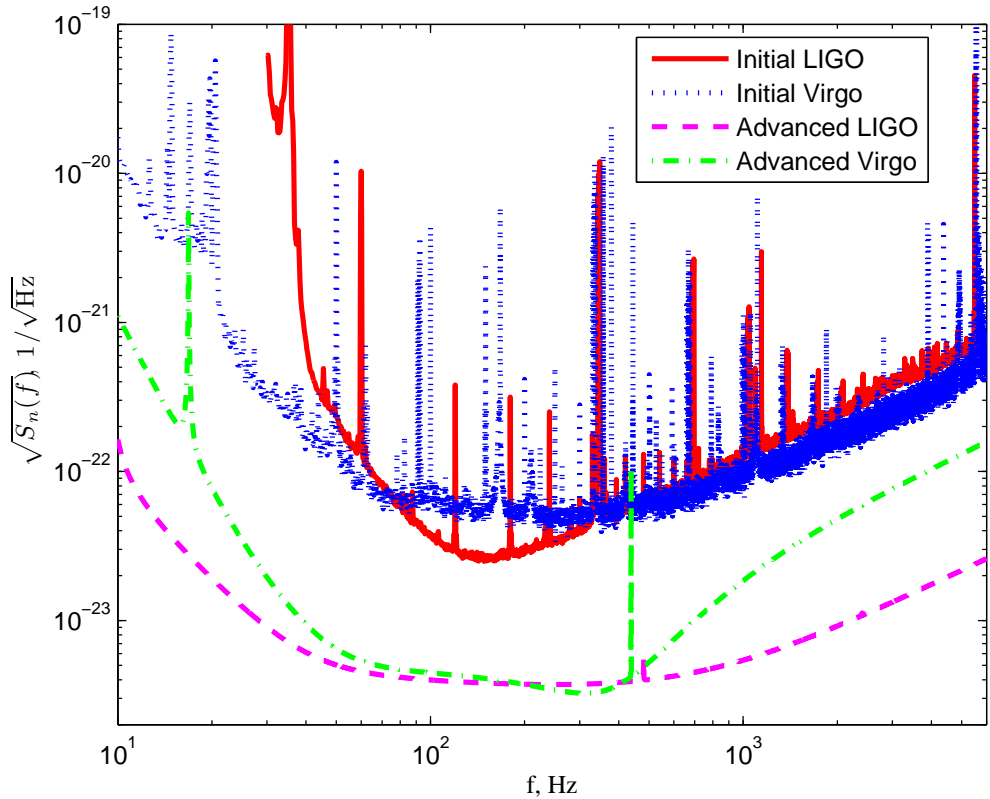}
\caption{Noise amplitude spectral densities (ASDs) as a function of frequency.  The Initial LIGO noise ASD (solid red curve) corresponds to the typical detector sensitivity as measured from data taken during the S5 run \cite{PSD:S5}.  The Advanced LIGO noise ASD (dashed magenta) represents a possible Advanced LIGO configuration with high laser power and zero detuning \cite{PSD:AL}.  The Initial Virgo noise ASD (dotted blue) was measured during Virgo's VSR2 run \cite{PSD:VSR2}.  The Advanced Virgo noise ASD (dash-dotted green) is based on the Advanced Virgo Baseline Design \cite{PSD:AV}.}
\label{fig:NoiseCurves}
\end{figure}

The Initial LIGO noise PSD $S_n(f)$ is based on the typical detector sensitivity as measured from data taken during the S5 run \cite{PSD:S5}.  Specifically, the noise spectrum corresponds to a time when the Hanford 4km detector operated near its S5-run mode for the $1.4$--$1.4$ $M_\odot$ inspiral horizon.   
The Advanced LIGO noise PSD is based on the zero detuning, high laser power configuration as described in the public LIGO document T0900288 \cite{PSD:AL}. This configuration, also known as Mode 1b, assumes 125 W input laser power, 20\% signal recycling mirror  (SRM) transmissivity, and no detuning of the signal recycling cavity.  
We emphasize that the Advanced LIGO noise PSD is merely a prediction, and actual noise PSDs may differ from it.  In particular Advanced LIGO has several possible configurations with different laser power, SRM transmissivity, and detuning of the signal recycling cavity, producing different PSDs.  
The LIGO noise amplitude spectral densities (ASDs) are plotted in Figure \ref{fig:NoiseCurves} together with the noise ASDs for Initial Virgo, as measured on October 20, 2009 during Virgo's Second Science Run \cite{PSD:VSR2}, and Advanced Virgo, as described in the Advanced Virgo Baseline Design document \cite{PSD:AV}.
We note that confusion noise from a background of unresolved CBCs is not expected to be a problem even for Advanced LIGO and Virgo detectors \cite{RegimbauHughes:2009}.

For the purpose of calculating the NS-NS, NS-BH, and BH-BH rates in Table \ref{detrates}, we assumed that all NSs had a mass of $1.4 M_\odot$ and all BHs had a mass of $10 M_\odot$ \footnote{These are different from the masses of $(1.35\pm0.04) M_\odot$ and $(5\pm1) M_\odot$ assumed for NSs and BHs, respectively, when placing upper limits on rates in LIGO result papers to date, e.g.~\cite{Abbott:2009qj}.}.  Although we know that neutron stars and black holes will cover a range of masses (see, e.g., \cite{Oshaughnessy:2008, MandelOShaughnessy:2010}), our knowledge of the mass distribution is not sufficient at present to warrant more detailed models, and the uncertainties in the coalescence rates dominate errors from the simplified assumptions about component masses.    For a single interferometer and a detection threshold of $\rho=8$, this assumption yields Initial LIGO $D_{\rm horizon}$ values 33 Mpc / 70 Mpc / 161 Mpc
for NS-NS / NS-BH / BH-BH searches, respectively.   
For Advanced  LIGO, the three $D_{\rm horizon}$ values are 445 Mpc / 927 Mpc / 2187 Mpc, respectively.  

Again, we reiterate that these horizon distances are computed using the noise PSD of a single interferometer and that the cosmological redshift is not included.  On average, if the noise is Gaussian and stationary and the search is optimal, $X$ detectors with the same noise power spectral density will increase the SNR and the horizon distance by a factor of $\sqrt{X}$ relative to a single detector for a fixed network detection threshold $\rho$.   On the other hand, detection pipelines which have to contend with non-Gaussian, non-stationary noise in the detectors may require higher SNR thresholds for detection in order to achieve desired false alarm rates than what was assumed here.  Thus, although we have used the noise PSDs of a single LIGO interferometer in the calculation, the detection rates are intended to approximate the performance of the LIGO-Virgo network.  The users of this document are encouraged to compute their own rates for different noise PSDs, typical masses, etc., by recomputing $D_{\rm horizon}$ and scaling the rates in Table \ref{detrates} by a factor of $N_G(D_{\rm  horizon})/N_G(D^0_{\rm horizon})$, or $(D_{\rm horizon}/D^0_{\rm horizon})^3$ for $D_{\rm horizon} \gtrsim 30\ {\rm Mpc}$.

\section{Derivation of Compact Binary Coalescence Rates\label{sourcetypes}}

\subsection{NS-NS rates\label{NSNS}}

\begin{table}[h!]
\caption{Estimates of NS-NS inspiral rates.\label{NSNStable}}
\begin{tabular}{c@{\quad\vline\quad}c@{\quad}c@{\quad}c@{\quad}c}
\hline
Rate model &  \Rlow & \Rre & \Rpl & \Rup\\ 
& MWEG$^{-1}$ Myr$^{-1}$ & MWEG$^{-1}$  Myr$^{-1}$ & MWEG$^{-1}$ Myr$^{-1}$ & MWEG$^{-1}$ Myr$^{-1}$  \\
\hline

Extrapolation: Model 6 of Kim et al.~\cite{Kim:2003kkl}\footnotemark[1]  & 16.9& 83.0 & 292.1 &\\
Extrapolation: Model 14 of Kim et al.~\cite{Kim:2003kkl}\footnotemark[1]  & 1.0& 3.8&13.2 &\\
Extrapolation: Model 15 of Kim et al.~\cite{Kim:2003kkl}\footnotemark[1]  & 43.1& 223.7&817.5 &\\
O'Shaughnessy et al.~pop.~synth.~\cite{Oshaughnessy:2008}\footnotemark[2]  & 5 & 30 & 300 & \\
Voss \& Tauris pop.~synth.~\cite{VossTauris:2003}\footnotemark[3] & 0.54 & 1.5 & 17 & \\
Belczynski et al.~pop.~synth.: model A of \cite{Belczynski:2007}\footnotemark[4] & & 12 & & \\
Belczynski et al.~pop.~synth.: model B of \cite{Belczynski:2007}\footnotemark[4] & & 7.6 & & \\
Belczynski et al.~pop.~synth.: model C of \cite{Belczynski:2007}\footnotemark[4] & & 68 & & \\
Nelemans pop.~synth.~\cite{Nelemans:2003}\footnotemark[5] & 0.5 & 25 & 1250 & \\
``Double-core'' scenario: Dewi et al.~\cite{Dewi:2006}\footnotemark[6] & 0.91 & 12.10 & &\\
With ellipticals: de Freitas Pacheco et al.~\cite{pacheco:2005}\footnotemark[7] & & 34 & &\\
Supernova Ib/Ic limit \cite{Kim:2006}\footnotemark[8] & & & & 3900\\
\hline
\end{tabular}

\footnotetext[1]{The estimates, based on an extrapolation from known 
NS-NS systems observed as binary pulsars, are taken from Table I of 
Kalogera et al.~\cite{Kalogera:2004tn}.  The model numbers refer to various pulsar 
luminosity distribution models (see Kim et al.~\cite{Kim:2003kkl} for an 
explanation).  Model 6 is the reference model: $L_{\rm min}=0.3$ mJy 
kpc$^2$, $p=2$.  Model 14 yields the lowest, and model 15 the highest rate estimates.  All values 
are based on additional assumptions about the pulsar lifetimes, beaming 
factors, etc., which could lead to significant systematic errors in the 
extrapolated rates.}

\footnotetext[2]{Predictions from constrained population-synthesis 
models \cite{Oshaughnessy:2008}.  A visual estimate of the center of the 
NS-NS probability distribution peak of Figure 6 is used as the value of 
\Rre; a visual estimate of the left and right edges of this peak are used as the 
values of \Rlow~and \Rpl.}

\footnotetext[3]{Predictions from the population-synthesis study of Voss \& Tauris \cite{VossTauris:2003}. 
The realistic estimate is taken from model A and the plausible pessimistic / optimistic rates are based on the lowest (model I) and highest (model B) predictions from Table 7 of \cite{VossTauris:2003}; the range may significantly underestimate the true uncertainty.}

\footnotetext[4]{Predictions from the population-synthesis studies of Belczynski et al.~\cite{Belczynski:2007}, which analyze the impact of assumptions about common-envelope evolution.  See section \ref{BHBH} for details regarding models A, B, and C.  Values are taken from Table 2 of \cite{Belczynski:2007}.}

\footnotetext[5]{Predictions from population-synthesis models of 
Nelemans \cite{Nelemans:2003}.  The realistic estimate is taken from the 
merger rate quoted in Table 1 of \cite{Nelemans:2003}.  The plausible 
pessimistic / optimistic estimates are obtained by dividing / multiplying that realistic estimate 
by the uncertainty factor of 50 quoted in that table.}

\footnotetext[6]{Predictions for NS-NS binaries that form through the ``double-core'' scenario.  The plausible pessimistic and realistic rates are taken to be the lowest and highest merger rates in Table 1 of Dewi et al.~\cite{Dewi:2006}.} 

\footnotetext[7]{This prediction from de Freitas Pacheco et al.~\cite{pacheco:2005} is the first to include the contribution of elliptical galaxies to CBC rates. In the absence of stated uncertainties in this contribution, the quoted mean local coalescence rate is taken as the realistic rate.}

\footnotetext[8]{The upper limit comes from the assumption that the 
formation of a NS-NS system requires a type Ib/c supernova \cite{Kim:2006}.  This upper 
limit is given as the mean SN Ib/c rate plus $2\sigma$, which is quoted 
as $1100\pm600$ per Myr per $L_{10}$ in \cite{Cappellaro:1999}.}



\end{table}

There are two distinct methods for estimating NS-NS merger rates.  The  first method is based on extrapolating from the observed sample of NS binaries  detected via pulsar measurements; the second method is based on population-synthesis codes, in which some of the unknown model parameters are 
constrained by observations and others are constrained by theoretical considerations.  We quote rate estimates from both of these methods in Table \ref{NSNStable}.

The most recent references for the first method, extrapolating double neutron-star merger rates from observed merging binary pulsars in the Galaxy, are those by Kalogera et al.~\cite{Kalogera:2004tn} and  Kim et al.~\cite{Kim:2006}.  These studies differ from previous work in statistically accounting for the small-number sample by using the Poisson distribution. This empirical method has significantly fewer free parameters than the population-synthesis models.  However, it does suffer from a small 
sample of observed merging NS-NS systems in the Galaxy\footnote{
The rates quoted in the top three lines of Table \ref{NSNStable} are based on extrapolation from three binary-pulsar systems: B1913+16, B1534+12, and J0737-3039.  Two additional field binary pulsars have been confirmed since these rates were computed in \cite{Kalogera:2004tn}: J1756-2251 and J1906+0746. According to Kim et al.~\cite{Kim:2006}, carefully including J1756-2251 in the rates estimates is non-trivial because of different selection effects of the acceleration search which found it, but should increase the rate predictions by only $\sim4\%$.  Meanwhile, the inclusion of J1906+0746, a relatively short-lived system, would increase the rates by almost a factor of $2$~\cite{Kim:2006}.  Therefore, the rates listed in Table \ref{NSNStable} may be a conservative estimate.}, 
and the implicit assumption that these form a good  representation of the total double neutron-star population.  Moreover,  as described in \cite{Kim:2003kkl}, the reconstruction of the Galactic  neutron star binary population relies on an understanding of the pulsar survey selection effects, and, therefore, on the pulsar luminosity distribution.  This distribution is described by two variables: the minimum pulsar luminosity $L_{\rm min}$ and the negative slope $p$ of the pulsar luminosity distribution power law.  Different choices of these variables, still consistent with the overall pulsar population observations, could change the merger rates by an order of magnitude.  An attempt has been made to fold in the distribution of $L_{\rm min}$ and $p$ into the rates calculation \cite{Kim:2006}, yielding a likely NS-NS merger rate of $13$ Myr$^{-1}$ MWEG$^{-1}$; however, this attempt suffered from out-of-date constraints on $L_{\rm min}$.  Therefore, the rates quoted here do not incorporate uncertainties in the pulsar luminosity function 
directly.  Instead, for the plausible optimistic / plausible pessimistic estimates, we quote rates at the upper / lower end of the $95\%$ confidence interval for the model yielding the highest / lowest rates from Table 1 of \cite{Kalogera:2004tn}.  For the likely estimate, we quote the rates at peak probability for the 
preferred pulsar luminosity distribution model 6 ($L_{\rm min}=0.3$ mJy kpc$^2$, $p=2$) from Table 1 of \cite{Kalogera:2004tn}, because this value of $L_{\rm min}$ is the lowest luminosity of all currently known pulsars.  Several additional assumptions are made which could systematically bias the rate estimates, e.g., the age of the pulsar at the time of detection, which could change rates by no more than a factor of $2$, and the beaming fraction, which has not yet been measured directly for all systems.

The most recent reference for the second method, in which binary merger rates are obtained from population-synthesis codes, is O'Shaughnessy et al.~\cite{Oshaughnessy:2008}.  Other population-synthesis studies carry out a much more limited exploration of the model parameter space and/or
do not quantitatively apply empirical constraints from pulsar observations. The population-synthesis code StarTrack  \cite{Belczynski:2008} used in \cite{Oshaughnessy:2008} contains a number of free parameters; seven of them have been selected as having the most significant impact on the model outcome and these are allowed to vary over a large range of values.  In the absence of firm empirical guidance, flat priors are used on these seven parameters, which include birth kick velocities, common-envelope efficiency, and the companion mass distribution.  However, narrower choices of these parameters could significantly impact rates: for example, if kicks are preferentially aligned with pre-supernova spin, NS-NS coalescence rates could decrease by up to a factor of five \cite{Postnov:2008}.
Additional constraints from observations are applied to select only those choices of models that are consistent with observations.  In \cite{Oshaughnessy:2008}, five such constraints are used: (i) the observed sample of merging binary pulsars; (ii) the observed sample of wide binary pulsars; (iii) the observed sample of white dwarf -- pulsar systems; (iv) the empirically derived Type II supernova rate; and (v) the empirically derived Type Ib/c supernova rate.  However, the supernova rates do not provide strong constraints (see Fig.~1 of  \cite{Oshaughnessy:2008}), so the meaningful constraints come from observed Galactic compact object binaries involving pulsars.   In this way, the rate estimates obtained via the second method after applying observational constraints are not truly independent from the estimates obtained via the first method, since the same observations are used.  Moreover, the second method suffers from some of the same uncertainties as the first method, particularly the need to estimate the pulsar luminosity distribution in order to reconstruct the Galactic pulsar population; the values $L_{\rm min}=0.3$ mJy kpc$^2$, $p=2$ are used in \cite{Oshaughnessy:2008}.  The rate  estimates are sensitive to the choice of observational constraints imposed on the parameters in the population-synthesis models; for instance, the rates changed by a factor of $\sim 5$ between \cite{Oshaughnessy:2005} and \cite{Oshaughnessy:2008} (other reasons for this difference include better accounting for systematic errors in the observations of wide NS-NS binaries and a more extensive coverage of the model parameter space).  On the other hand, the population-synthesis method is the only one available for estimating NS-BH and BH-BH rates (see below), since they have not been observed electromagnetically.  For the optimistic / pessimistic Galactic rates we estimate the location of the right / left edge of the rate probability distribution peak in Figure 6 of \cite{Oshaughnessy:2008}, while for the likely rate we take the location of the maximum.

The binary evolution models in the population-synthesis study of Voss \& Tauris \cite{VossTauris:2003} differ in several significant ways from the models of O'Shaughnessy et al.; for instance, they do not allow for hypercritical accretion to occur during a common envelope phase.  They qualitatively confirmed that their models match the observed Galactic binary-pulsar distribution.  Voss \& Tauris investigate their models by varying one astrophysical parameter at a time, which limits parameter-space coverage and makes it difficult to estimate the range of uncertainty in rate predictions.  For the realistic NS-NS Galactic merger rate, we used the value of their default model A in Table 7 (see also Table 4); for the plausible pessimistic / optimistic rates, we used the lowest (model I) and highest (model B) predictions from Table 7 of \cite{VossTauris:2003}, respectively, with the understanding that since only one parameter was varied at a time, this may significantly underestimate the true uncertainty.

Belczynski et al.~\cite{Belczynski:2007} specifically examined how the treatment of common-envelope evolution affects CBC rate predictions.  These effects are most striking for BH-BH rates, and are therefore discussed in more detail in Section \ref{BHBH}.  The authors report results for three specific models, A, B, and C, in Table 2 of \cite{Belczynski:2007}, which we list as realistic rates here, since quantitative estimates of uncertainties in the rate predictions are not available.

Rate estimates obtained by Nelemans and collaborators on the basis of population-synthesis studies \cite{Nelemans:2001, Nelemans:2003} are also included.  The likely estimate is taken directly from the ``merger rate'' column of Table 1 of \cite{Nelemans:2003}, while plausible optimistic / pessimistic estimates are obtained by multiplying / dividing this value by the uncertainty factor of $50$ quoted for 
NS-NS binaries in the same table.  These estimates, however, do not  include model constraints on the basis of pulsar observations.

Dewi et al.~\cite{Dewi:2006} model the ``double-core'' scenario, in which two nearly equal-mass stars form an NS-NS binary through a double common-envelope phase which they enter after both stars have already evolved off the main sequence (see also Brown \cite{1995ApJ...440..270B}).  The authors find that even limiting their attention to NS-NS binaries formed from two helium stars in close orbit yields merger rates between $0.91$ and $12.10$ per MWEG per Myr, depending on assumptions about mass transfer and common-envelope efficiency.  Since other NS-NS formation scenarios are not included, we use these values, taken from Table 1 of \cite{Dewi:2006}, as the plausible pessimistic and realistic merger rates.

The first effort to include the contribution of elliptical galaxies to the detection rates was undertaken by de Freitas Pacheco et al.~\cite{pacheco:2005}.  The authors used a mixture of population-synthesis techniques and fitting to observational data on Galactic pulsars; they calibrated their Galactic merger rate with a sophisticated model of the star formation history of the Galaxy.  The authors found that the inclusion of elliptical galaxies, which have little present-day star formation but could still contribute to CBC rates through delayed mergers, approximately doubled the local (to redshift $z < 0.01$) merger rate, to $34$ NS-NS mergers per Myr.  The uncertainty in the merger rate in ellipticals was not explored, so we list the quoted mean local coalescence rate of $34$ per Myr as the realistic rate prediction.

We note that all realistic rate estimates quoted above fall in the range set by the top three lines of Table \ref{NSNStable}, which is the range used in the Executive Summary.

Since current theoretical understanding predicts that the second neutron star in an NS-NS system should be born in a Type Ib/Ic supernova, the rate of such supernovae provide an upper limit on NS-NS rates \cite{Kim:2006}.  We quote the upper limit as the mean SN Ib/c rate plus $2 \sigma$ from Table 4 of \cite{Cappellaro:1999}, taking care to convert from a rate per $L_{10}$ to a rate per MWEG.  Note that population-synthesis codes predict that only ${\cal O}(5\%)$ of all SN Ib/c are involved in the formation of NS-NS systems \cite{Belczynski:2008}.

Dynamical interactions in globular clusters are not expected to contribute significantly to the total rate of NS-NS coalescences \cite{Sadowski, Ivanova:2008} and are not included in the results in this Section. For example, Grindlay et al.~\cite{Grindlay:2006} estimate the combined NS-NS merger rate in all globular clusters in the Galaxy at $40$ per Gyr, which is three orders of magnitude less than the predictions for field mergers.  Additional discussion of the contribution of dynamical interactions to rates for binaries containing neutron stars can be found in Section \ref{GRB}, while the contribution of dense stellar environments to BH-BH rates is described in Section \ref{BHBHdyn}.



\subsection{NS-BH rates\label{NSBH}}

\begin{table}[h!]
\caption{Estimates of NS-BH inspiral rates.\label{NSBHtable}}
\begin{tabular}{c@{\quad\vline\quad}c@{\quad}c@{\quad}c@{\quad}c}
\hline
Rate model & \Rlow & \Rre & \Rpl & \Rup\\ 
& MWEG$^{-1}$ Myr$^{-1}$ & MWEG$^{-1}$ Myr$^{-1}$ & MWEG$^{-1}$ Myr$^{-1}$ & MWEG$^{-1}$ Myr$^{-1}$ \\
\hline

O'Shaughnessy et al.~pop.~synth.~\cite{Oshaughnessy:2008}\footnotemark[1] & 0.05 & 3 & 
100 & \\
Voss \& Tauris pop.~synth.~\cite{VossTauris:2003}\footnotemark[2] & 0.2 & 0.58 & 5 & \\
Belczynski et al.~pop.~synth.: model A of \cite{Belczynski:2007}\footnotemark[3] & & 0.07 & & \\
Belczynski et al.~pop.~synth.: model B of \cite{Belczynski:2007}\footnotemark[3] & & 0.09 & & \\
Belczynski et al.~pop.~synth.: model C of \cite{Belczynski:2007}\footnotemark[3] & & 3.2 & & \\
Nelemans pop.~synth.~\cite{Nelemans:2003}\footnotemark[4] & 0.2 & 10 & 500 & \\
``Double-core'' scenario: Dewi et al.~\cite{Dewi:2006}\footnotemark[5] & 0.14 & 6.32 & &\\
\hline
\end{tabular}

\footnotetext[1]{Predictions from constrained population-synthesis 
models \cite{Oshaughnessy:2008}.  A visual estimate of the center of the 
NS-BH probability distribution peak of Figure 6 is used as the value of 
\Rre; a visual estimate of the left / right edge of this peak is used as the 
values of \Rlow / \Rpl.}

\footnotetext[2]{Predictions from the population-synthesis study of Voss \& Tauris \cite{VossTauris:2003}. 
The realistic estimate is taken from model A and the plausible pessimistic / optimistic rates are based on the lowest (model D) and highest (model B) predictions from Table 7 of \cite{VossTauris:2003}.  The values for BHNS and NSBH rates are summed.  The range may significantly underestimate the true uncertainty.}

\footnotetext[3]{Predictions from the population-synthesis studies of Belczynski et al.~\cite{Belczynski:2007}, which analyze the impact of assumptions about common-envelope evolution.  See section \ref{BHBH} for details regarding models A, B, and C.  Values are taken from Table 2 of \cite{Belczynski:2007}.}

\footnotetext[4]{Predictions from population-synthesis models of 
Nelemans \cite{Nelemans:2003}.  The realistic estimate is taken from the
merger rate quoted in Table 1 of \cite{Nelemans:2003}.  The plausible
pessimistic and optimistic estimates are obtained, respectively, by dividing and multiplying 
that realistic estimate by the uncertainty factor of 50 quoted in that table.}

\footnotetext[5]{Predictions for NS-BH binaries that form through the ``double-core'' scenario.  The plausible pessimistic and realistic rates are taken to be the lowest and highest merger rates in Table 2 of Dewi et al.~\cite{Dewi:2006}.} 


\end{table}

Because of the lack of observations of coalescing compact-object binaries containing black holes, NS-BH rates can only be based on predictions from population-synthesis models, as discussed in section \ref{NSNS}.  

The O'Shaughnessy et al.~rates that we quote are based on the same constrained population-synthesis models as the NS-NS rates described earlier.   For the plausible pessimistic / optimistic Galactic rates we estimate the location of the left  / right edge of the rate probability distribution peak in Figure 6 of \cite{Oshaughnessy:2008}, while for the likely rate we take the location of the PDF maximum.

Voss \& Tauris population-synthesis results are taken from Table 7 of \cite{VossTauris:2003}.  Voss \& Tauris differentiate BHNS and NSBH merger rates based on which binary component was the first to evolve; however, we add their BHNS and NSBH values for this document.  For the realistic NS-BH Galactic merger rate, we used the value of their default model A in Table 7 (see also Table 4); for the plausible pessimistic / optimistic rates, we used the lowest (model D) and highest (model 
B) predictions from Table 7 of \cite{VossTauris:2003}, respectively, with the understanding that since only one parameter was varied at a time, this may significantly underestimate the true uncertainty.

Belczynski et al.~\cite{Belczynski:2007} have examined the effect of the treatment of common-envelope evolution, as described in more detail in section \ref{BHBH}.  The authors report results for three specific models, A, B, and C, in Table 2 of \cite{Belczynski:2007}, which we list as realistic rates here, since quantitative uncertainties in the rate predictions are not available.

Rate estimates obtained by Nelemans and collaborators via population-synthesis studies \cite{Nelemans:2001, Nelemans:2003} are also included.  The  ``merger rate'' column of Table 1 of
\cite{Nelemans:2003} is used for the likely estimate.  We divide/multiply this value by the uncertainty factor of $50$ quoted for this source type in the same table to obtain the plausible pessimistic / optimistic estimates.  These estimates, however, do not include model constraints on the basis of empirical observations.

Dewi et al.~\cite{Dewi:2006} applied their ``double-core'' scenario (see Section \ref{NSNS}) to the formation of NS-BH binaries, in which the carbon-oxygen core of the primary collapses to form a black hole after the contact phase.  Depending on assumptions about common-envelope efficiency, final black-hole mass, and black-hole birth kick velocity, they found NS-BH merger rates between $0.14$ and $6.32$ per MWEG per Myr.  Since other NS-BH binary formation scenarios are not included, we use these values, taken from Table 2 of \cite{Dewi:2006}, as the plausible pessimistic and realistic merger rates.

All realistic estimates quoted above fall in the range set by \cite{Oshaughnessy:2008} (see 
top line of Table \ref{NSBHtable}), which is the range used in the Executive Summary.

We note that there have been a number of studies of binaries composed of a black hole and a recycled pulsar (e.g., \cite{Sipior:2004, Pfahl:2005}).  However, since such systems likely form only a small subset of all NS-BH systems, where the NS may or may not be a recycled pulsar \cite{Pfahl:2005}, we do not include them here.


\subsubsection{Short gamma-ray bursts \label{GRB}}
Recently, several authors have attempted to extract rates for NS-NS and NS-BH coalescences from the rates of observed short hard gamma-ray bursts (SGRBs).  In particular, some models favor dynamical formation scenarios in globular clusters for at least some of the SGRB progenitors \cite{Grindlay:2006, Hopman:2006}.  According to some estimates, extrapolating the rates of SGRBs could yield higher overall rates for NS-NS and NS-BH coalescences than those described above \cite{Guetta:2009, Dietz:2009}.  However, most simulations indicate that dynamical effects are not a significant contribution to coalescence rates for binaries containing neutron stars, largely because more massive black holes are expected to sink to the centers of dense stellar environments and substitute into binaries during dynamical interactions \cite{Sadowski}.  Extrapolations from GRB measurements suffer from many uncertainties regarding the selection biases in SGRB observations, such as the unknown beaming fraction of SGRBs \cite{nakar07}, which may be different for binaries formed in clusters and in the field \cite{Grindlay:2006}.  Additionally, such estimates rely on the assumption that all SGRBs arise from coalescences of NS-NS or NS-BH systems following inspirals driven by gravitational-wave emission; however, this is not the only possible formation scenario for SGRBs (see \cite{Virgili:2009, Lee:2009} for some of the suggested alternatives).   We also note that, given the current observational and theoretical uncertainties, the observed SGRB rates broadly agree with predictions of NS-NS and NS-BH merger rates from isolated binary evolution alone \cite{OShaughnessy:2008grb, Belczynski:2008grb}. In view of the above, we choose to not include rates extrapolated from SGRB observations at this time; however, this approach is a very promising one, and could yield interesting CBC rate estimates once SGRB formation channels are well-understood and selection effects are accounted for.

\subsection{BH-BH rates\label{BHBH}}

\begin{table}[h!]
\caption{Estimates of BH-BH inspiral rates.\label{BHBHtable}}
\begin{tabular}{c@{\quad\vline\quad}c@{\quad}c@{\quad}c@{\quad}c@{\quad}c}
\hline
Rate model & \Rlow & \Rre & \Rpl & \Rup\\ 
\hline
O'Shaughnessy et al.~pop.~synth.~\cite{2007PhR...442...75K}\footnotemark[1] (MWEG$^{-1}$ Myr$^{-1}$)  & 0.01 & 0.4 & 
30 & \\
Voss \& Tauris pop.~synth.~\cite{VossTauris:2003}\footnotemark[2]  (MWEG$^{-1}$ Myr$^{-1}$) & 1.3 & 9.7 & 76 & \\
Belczynski et al.~pop.~synth.: model A of \cite{Belczynski:2007}\footnotemark[3] (MWEG$^{-1}$ Myr$^{-1}$)& & 0.02 & & \\
Belczynski et al.~pop.~synth.: model B of \cite{Belczynski:2007}\footnotemark[3] (MWEG$^{-1}$ Myr$^{-1}$)& & 0.01 & & \\
Belczynski et al.~pop.~synth.: model C of \cite{Belczynski:2007}\footnotemark[3] (MWEG$^{-1}$ Myr$^{-1}$)& & 7.7 & & \\
Nelemans pop.~synth.~\cite{Nelemans:2003}\footnotemark[4] (MWEG$^{-1}$ Myr$^{-1}$) & 0.1 & 5 & 250 & \\
``Double-core'' scenario: Dewi et al.~\cite{Dewi:2006}\footnotemark[5] (MWEG$^{-1}$ Myr$^{-1}$) & 0.19 & 19.87 & &\\
Globular cluster dynamics \cite{OLeary:2007}\footnotemark[6] (Mpc$^{-3}$ Myr$^{-1}$) & $10^{-4}$ & 0.05  & & 1\\
Globular cluster dynamics and pop.~synth.~\cite{Sadowski}\footnotemark[7] (GC$^{-1}$ Gyr$^{-1}$) & & 2.5 & & \\
Nuclear cluster w/ MBH  \cite{OLeary:2008}\footnotemark[8] (NC$^{-1}$ Myr$^{-1}$) & $2\times10^{-4}$ & $1.3\times10^{-3}$ & 0.015 & \\
Nuclear cluster w/out MBH \cite{MillerLauburg:2008}\footnotemark[9] (NC$^{-1}$ Myr$^{-1}$)  & & 0.3 & &\\
\hline
\end{tabular}

\footnotetext[1]{Predictions from constrained population-synthesis 
models \cite{2007PhR...442...75K}.  A visual estimate of the center of the 
BH-BH probability distribution peak in the panel of Figure 15 is used as the value of 
\Rre; visual estimate of the left / right edges of this peak are used as the 
values of \Rmin / \Rpl.}

\footnotetext[2]{Predictions from the population-synthesis study of Voss \& Tauris \cite{VossTauris:2003}. 
The realistic estimate is taken from model A and the plausible pessimistic / optimistic rates are based on the lowest (model D) and highest (model B) predictions from Table 7 of \cite{VossTauris:2003}.  The range may significantly underestimate the true uncertainty.}

\footnotetext[3]{Predictions from the population-synthesis studies of Belczynski et al.~\cite{Belczynski:2007}, which analyze the impact of assumptions about common-envelope evolution.  See below for details regarding their models A, B, and C.  Values are taken from Table 2 of \cite{Belczynski:2007}.}

\footnotetext[4]{Predictions from population-synthesis models of Nelemans \cite{Nelemans:2003}.  The realistic estimate is taken from the merger rate quoted in Table 1 of \cite{Nelemans:2003}.  The plausible
pessimistic / optimistic estimates are obtained by dividing / multiplying that realistic estimate by the uncertainty factor of 50 quoted in that table.}

\footnotetext[5]{Predictions for BH-BH binaries that form through the ``double-core'' scenario.  The plausible pessimistic and realistic rates are taken to be the lowest and highest merger rates in Table 2 of Dewi et al.~\cite{Dewi:2006}.} 

\footnotetext[6]{Predictions for BH-BH merger rates in dense BH subclusters at the cores of stellar clusters \cite{OLeary:2007}.  The predicted rates are $\approx g_{\rm evap}\ g_{\rm cl}$ Mpc$^{-3}$ Myr$^{-1}$, where $g_{\rm evap}\ g_{\rm cl}$ should be larger than $10^{-4}$ (plausible pessimistic value), is likely $5\times10^{-2}$ (realistic value), and could be as high as $1$ (upper limit).}

\footnotetext[7]{Predictions for BH-BH merger rates in globular cluster cores in thermal equilibrium \cite{Sadowski}.  The predicted rate for a globular cluster of mass $4.8\times 10^5\ M_\odot$ is $2.5$ BH-BH coalescences per Gyr, according to Section 3.3.}

\footnotetext[8]{Predictions from models of 2-body BH-BH dynamical scattering in galactic nuclei \cite{OLeary:2008}.  The plausible pessimistic, realistic, and plausible optimistic rates per nuclear cluster are taken from models A$\beta$3, E2, and F1 of Table 1 of \cite{OLeary:2008}.}

\footnotetext[9]{Predictions from models of nuclear clusters of small galaxies without massive black holes \cite{MillerLauburg:2008}.  The realistic rate is quoted based on the prediction of a ``merger rate of  $>\, 0.1 \times {\rm a\ few}$'' per Myr per galaxy (see Section 3 of  \cite{MillerLauburg:2008}).}

\end{table}

There are two distinct scenarios for the formation of double black-hole binaries close enough to coalesce through gravitational-wave emission.  The first is the isolated binary-evolution scenario, which is expected to be the dominant scenario for NS-NS and NS-BH systems described above.  The second scenario, which can be significant for BH-BH systems because of their higher mass, is the dynamical-formation scenario, in which dynamical interactions in dense stellar environments play a significant role in forming and/or hardening the black-hole binary before coalescence driven by radiation reaction.  This scenario can be particularly important in globular clusters \cite{OLeary:2007, Sadowski} and nuclear star clusters with \cite{OLeary:2008} or without \cite{MillerLauburg:2008} a massive black hole; however, because of the uncertainties involved in the dynamical-formation scenario predictions, and the difficulty of assigning ranges given the limited number of models considered thus far, we do not currently include these predictions in the Executive Summary tables (see Section \ref{BHBHdyn}).  

\subsubsection{BH-BH rates via the isolated binary-evolution scenario}
Because of the lack of observations of coalescing binaries containing black holes, BH-BH rates can only be based on predictions from population-synthesis models, constrained as discussed in section 
\ref{NSNS}.  The most recent published constrained population-synthesis results \cite{Oshaughnessy:2008} do not include BH-BH rates, because BH-BH mergers can be significantly delayed relative to binary formation, so that elliptical galaxies with little current star formation and low blue-light luminosities can contribute significantly to BH-BH rates.  In the meantime, we use results from \cite{2007PhR...442...75K}, which do not properly account for the delay between star formation and merger.  For the plausible pessimistic / optimistic Galactic rates we estimate the location of the left / right edge of the rate  probability distribution peak in the top panel of Figure 15 of \cite{2007PhR...442...75K},  while for the likely rate we take the location of the center of the peak.  Note that an older population-synthesis study \cite{Oshaughnessy:2005} only applies the observed Galactic double neutron-star population as a constraint and does not properly include systematics for wide NS-NS binaries.   

Voss \& Tauris population-synthesis results are taken from Table 7 of \cite{VossTauris:2003}.   For the realistic BH-BH Galactic merger rate, we used the value of their default model A in Table 7 (see also Table 4); for the plausible pessimistic / optimistic rates, we used the lowest (model D) and highest (model B) predictions from Table 7 of \cite{VossTauris:2003}, respectively, with the understanding that since only one parameter was varied at a time, this may significantly underestimate the true uncertainty.

Belczynski et al.~\cite{Belczynski:2007} find that many potential BH-BH progenitors enter a common-envelope phase while the donor star is evolving through the Hertzsprung gap.  Contrary to earlier studies (e.g., \cite{Belczynski:2002}), they indicate that such systems may very likely merge in the common-envelope phase, thereby inhibiting the formation of tight compact-object binaries.  They also find that accretion during the common-envelope phase should lead to considerably smaller mass gain.  The combined effect of these changes may strongly suppress the merger rates for BH-BH systems, and somewhat lowers other CBC rates.  Model A is the default model which includes both new effects; model B allows for the full hypercritical accretion during the common envelope phase, but still assumes that entering the common envelope while the donor crosses the Hertzsprung gap leads to merger; and model C is the model that does not include either effect and is closest to \cite{Belczynski:2002}.  Quantitative uncertainties in the rate predictions are not available, so we list all three models as the realistic rates here, taking the lower value (corresponding to normalization by the the star formation rate) from Table 2 of \cite{Belczynski:2007}.

Rate estimates obtained by Nelemans and collaborators via population-synthesis studies \cite{Nelemans:2003} are also included.  The likely estimate is taken directly from the ``merger rate'' column of Table 1 of \cite{Nelemans:2003}, while plausible pessimistic / optimisitic estimates are obtained by dividing / multiplying this value by the uncertainty factor of $50$ quoted for BH-BH binaries in the same table.  These estimates, however, do not  include model constraints on the basis of empirical observations.

Dewi et al.~\cite{Dewi:2006} applied their ``double-core'' scenario (see Section \ref{NSNS}) to the formation of BH-BH binaries.  Depending on assumptions about common-envelope efficiency, final black-hole mass, and black-hole birth kick velocity, they found BH-BH merger rates between $0.19$ and $19.87$ per MWEG per Myr.  Since other BH-BH binary formation scenarios are not included, we use these values, taken from Table 2 of \cite{Dewi:2006}, as the plausible pessimistic and realistic merger rates.

Again, all realistic BH-BH merger rate estimates quoted for the isolated binary evolution scenario 
fall in the range set by \cite{2007PhR...442...75K} (see top line of Table \ref{BHBHtable}), which is the range used in the Executive Summary.

\subsubsection{BH-BH rates via the dynamical-formation scenario \label{BHBHdyn}}

O'Leary, O'Shaughnessy and Rasio \cite{OLeary:2007} use N-body simulations to analyze the dynamics leading to BH-BH mergers in globular clusters.  Building on the results of \cite{OLeary:2006}, they consider a number of models of stellar clusters that differ in the assumptions about cluster histories, star formation, velocity dispersion and other properties to compute present-day merger rates of $\approx g_{\rm evap}\ g_{\rm cl}$ Mpc$^{-3}$ Myr$^{-1}$, where $g_{\rm cl}$ is the fraction of total star formation that occurs in clusters and $g_{\rm evap}$ is the fraction of all cluster-forming mass that possesses the birth conditions necessary to lead to the formation of a dense black-hole subcluster through gravitational segregation and cluster evaporation.  Based on weak constraints from globular-cluster observations, the authors argue that the fraction $g_{\rm evap}\ g_{\rm cl}$ should be larger than $10^{-4}$ (plausible pessimistic value), is likely $5\times10^{-2}$ (likely value), and could be as high as $1$ (upper limit).  The plausible optimistic value yields one event every two years with Initial LIGO.

Sadowski et al.~\cite{Sadowski} use a combination of a Monte Carlo code for dynamical interactions and the StarTrack code for stellar evolution to estimate the BH-BH merger rates in a globular cluster.  Unlike O'Leary et al.~\cite{OLeary:2007}, they assume that the black holes at the core of the cluster do not decouple into a subcluster, but remain in thermal equilibrium with other stars in the core and continuously interact with them through binary-single and binary-binary encounters.  The authors find that if the fraction of stellar mass initially contained in clusters (relative to the mass of stars in the field) is significant yet plausible, then the rate of dynamical BH-BH binary formation in clusters may exceed the rate of BH-BH binary formation through isolated binary evolution in the field.  They find rates of $2.5$ BH-BH coalescences per Gyr for globular clusters of mass $4.8\times 10^5\ M_\odot$.  Depending on the mass fraction in clusters, they conclude that overall Initial LIGO detection rates could range from $0.01$ to $1$, and Advanced LIGO rates could range from $25$ to $3000$ detections per year.  However, the authors have evolved only five clusters with identical choices for other parameters (e.g., a low metallicity $Z=0.001$), so that it is difficult to estimate the uncertainties in their predictions and determine a plausible range.

O'Leary, Kocsis and Loeb argue in \cite{OLeary:2008} that stellar-mass black holes in galactic nuclei with a supermassive black hole can create steep density cusps with enough scattering interactions to form a significant number of tight BH-BH binaries through direct 2-body scattering.  Because these are initially hyperbolic encounters that lead to capture through energy loss during the first periapsis passage, these binaries have the distinguishing feature of being eccentric; the binaries then coalesce on a timescale of hours.  The plausible pessimistic, likely, and plausible optimistic rates per nuclear cluster are taken from models A$\beta$3, E2, and F1 of Table 1, as the lowest, intermediate, and highest rates reported in that table.  These may be based on optimistic assumptions regarding the fraction of black holes in galactic nuclei and the extrapolation of the number density of galaxies to low masses.
Note that these are average rates per galactic center, not per MWEG; the authors extrapolate the distribution of massive black holes to $10^4\, M_\odot$ to obtain predictions for Advanced LIGO of 1 to 1000 detections per year, based on optimistic assumptions about the Advanced LIGO detection thresholds.

Meanwhile, Miller and Lauburg consider nuclear clusters of small galaxies that do not have massive black holes as possible sources of BH-BH coalescences \cite{MillerLauburg:2008}.  In these environments, the tightening of BH-BH binaries is driven primarily by 3-body interactions, with eventual inspiral due to radiation reaction.  Although rates for these processes, as well as for the two processes discussed above, depend heavily on the poorly-constrained mass function of black holes in the dense cores of clusters, the authors argue for a ``merger rate of  $>\, 0.1 \times {\rm a\ few}$'' per Myr per galaxy (see Section 3 of  \cite{MillerLauburg:2008}), so we quote 0.3 per Myr per nuclear cluster as the likely rate estimate.  Miller and Lauburg translate this rate into a prediction of several tens of detectable BH-BH inspiral events per year with Advanced LIGO.

Because of the uncertainties involved in the dynamical-formation scenario predictions, and the difficulty of assigning ranges given the limited number of models considered thus far, we do not currently include these predictions in the executive summary tables.  However, as can be seen from the preceding table, the dynamical-formation scenario could significantly increase the rates for BH-BH coalescences, particularly if the actual isolated binary-evolution rates fall on the low side of the predicted range while the dynamical rates are closer to the claimed upper limits of the range.    As additional confidence is gained through improved analytical understanding and numerical modeling, the dynamical-formation rates will, of course, need to be included in the overall BH-BH rate predictions.

\subsection{Rates of IMRIs into IMBHs}

\begin{table}[h!]
\caption{Rate estimates for intermediate-mass-ratio inspirals into 
intermediate-mass black holes.}
\begin{tabular}{c@{\quad\vline\quad}c@{\quad}c@{\quad}c@{\quad}c@{\quad}c}
\hline
Rate model & \Rlow & \Rre & \Rpl & \Rup\\ 
& GC$^{-1}$ Gyr$^{-1}$ & GC$^{-1}$ Gyr$^{-1}$ 
& GC$^{-1}$ Gyr$^{-1}$ & GC$^{-1}$ Gyr$^{-1}$  \\
\hline
Mandel et al., NS-IMBH IMRI  &  &  & 3\footnotemark[1] & 20\footnotemark[3]\\
Mandel et al., BH-IMBH IMRI  &  &  & 5\footnotemark[2] & 3\footnotemark[3]\\
\hline
\end{tabular}

\footnotetext[1]{The rate for inspirals of $1.4M_\odot$ NSs into a 
$100M_\odot$ IMBH via three-body hardening (Section 2.1 of 
\cite{Mandel:2007rates}).}

\footnotetext[2]{The rate for inspirals of $10M_\odot$ BHs into a 
$100M_\odot$ IMBH via three-body hardening (Section 2.1 of 
\cite{Mandel:2007rates}).}

\footnotetext[3]{Upper limit based on the growth of an IMBH by 
$300M_\odot$ in $10^{10}$ years exclusively through IMRIs of 
$1.4M_\odot$ NSs or $10M_\odot$ BHs (Section 3.3 of 
\cite{Mandel:2007rates}).}

\end{table}

The very existence of intermediate-mass black holes is still debatable \cite{Miller:2009}, so intermediate-mass-ratio inspirals (IMRIs) into IMBHs are an uncertain class of sources for the LIGO-Virgo network.  However, as described in \cite{Mandel:2007rates}, IMRIs of NSs or BHs into IMBHs in globular clusters could, under optimistic conditions, present an interesting Advanced LIGO-Virgo source.  (They are not likely to be a significant source for the Initial LIGO-Virgo network, both because of its lower detection range and higher low-frequency cutoff.)

The upper limits are obtained by assuming that most of the IMBH mass in a globular cluster comes from minor mergers that are potentially detectable  as IMRIs in the LIGO-Virgo band.  The plausible optimistic estimates are obtained by considering the timescales of binary formation, subsequent binary tightening through three-body interactions, and merger through radiation reaction from the emission of gravitational waves. However, even these optimistic rates are highly uncertain; for example, the fraction of globular clusters containing an IMBH of a suitable mass range ($\sim 50$ to a few hundred solar masses for Advanced LIGO-Virgo) is assumed to be $10\%$ without justification \cite{Mandel:2007rates}.  Detection rates for Initial and Advanced LIGO in Table \ref{detrates} are quoted directly from \cite{Mandel:2007rates}; a detection SNR threshold of 8 was assumed.  These rates assume that the IMBH mass is $\sim 100\ M_\odot$.  While IMRIs into IMBHs more massive than $\sim 400\ M_\odot$ will be outside the Advanced LIGO-Virgo frequency band, the ringdowns following such coalescences may be detectable (see Appendix B of \cite{Mandel:2007rates}).

\subsection{IMBH-IMBH rates}

\begin{table}[h!]
\caption{Estimates of IMBH-IMBH coalescence rates.}
\begin{tabular}{c@{\quad\vline\quad}c@{\quad}c@{\quad}c@{\quad}c}
\hline
Rate model & \Rlow & \Rre & \Rpl & \Rup\\ 
& GC$^{-1}$ Gyr$^{-1}$ & GC$^{-1}$ Gyr$^{-1}$ 
& GC$^{-1}$ Gyr$^{-1}$ & GC$^{-1}$ Gyr$^{-1}$\\
\hline
Fregeau et al.~& & & 0.007\footnotemark[1] & 0.07\footnotemark[2]\\ 
\hline
\end{tabular}
\footnotetext[1]{Assumes that $10\%$ of star clusters are sufficiently massive and have a sufficient binary fraction to form an IMBH-IMBH binary once in their lifetime, taken to be 13.8 Gyr \cite{imbhlisa-2006}.}
\footnotetext[2]{Assumes that all star clusters are sufficiently massive and have a sufficient binary fraction to form an IMBH-IMBH binary once in their lifetime, taken to be 13.8 Gyr \cite{imbhlisa-2006}.}
\end{table}

As mentioned above, the existence of intermediate-mass black holes is still uncertain, as is their prevalence and mass distribution if they do exist.  If the binary fraction in a young dense cluster exceeds $\sim 10\%$, and the deep core collapse timescale is shorter than $\sim 3$ Myr, IMBH-IMBH binaries could form via collisional runaway in young dense stellar clusters \cite{imbhlisa-2006}.  For IMBH masses considered in \cite{imbhlisa-2006}, the inspiral frequency would be too low for the inspiral to be detectable; however,  the LIGO-Virgo network could detect the merger and ringdown waveforms.  The fraction of clusters with a sufficient mass and binary fraction is scaled to 10\% without justification in \cite{imbhlisa-2006}; it obviously cannot exceed 1.  Because of the uncertainties involved, these results are listed as plausible optimistic estimates and upper limits.  Detection rates for Initial and Advanced LIGO in Table \ref{detrates} are quoted directly from \cite{imbhlisa-2006}; a detection SNR threshold of 8 was assumed.

Another proposed mechanism for forming IMBH-IMBH binaries is the collision of two globular clusters, each of which contains an IMBH \cite{Amaro:2006imbh}.  No LIGO-Virgo detection rates are provided in 
\cite{Amaro:2006imbh}.

\section*{Acknowledgements}
The authors gratefully acknowledge the support of the United States
National Science Foundation for the construction and operation of the
LIGO Laboratory, the Science and Technology Facilities Council of the
United Kingdom, the Max-Planck-Society, and the State of
Niedersachsen/Germany for support of the construction and operation of
the GEO600 detector, and the Italian Istituto Nazionale di Fisica
Nucleare and the French Centre National de la Recherche Scientifique
for the construction and operation of the Virgo detector. The authors
also gratefully acknowledge the support of the research by these
agencies and by the Australian Research Council, the Council of
Scientific and Industrial Research of India, the Istituto Nazionale di
Fisica Nucleare of Italy, the Spanish Ministerio de Educaci\'on y
Ciencia, the Conselleria d'Economia Hisenda i Innovaci\'o of the
Govern de les Illes Balears, the Foundation for Fundamental Research
on Matter supported by the Netherlands Organisation for Scientific Research, 
the Polish Ministry of Science and Higher Education, the FOCUS
Programme of Foundation for Polish Science,
the Royal Society, the Scottish Funding Council, the
Scottish Universities Physics Alliance, The National Aeronautics and
Space Administration, the Carnegie Trust, the Leverhulme Trust, the
David and Lucile Packard Foundation, the Research Corporation, and
the Alfred P. Sloan Foundation.
One of us (CK) would like to acknowledge the European Commission.




\bibliographystyle{unsrt}
\bibliography{../bibtex/iulpapers}
\end{document}